\documentclass[12pt]{iopart}
\usepackage{iopams}

\usepackage{graphicx}
\usepackage{dcolumn}
\usepackage{bm}

\newcommand{\bfr}{{\mathbf{r}}}
\newcommand{\bfR}{{\mathbf{R}}}
\newcommand{\ee}{{\mathrm{e}}}

\begin{document}

\title[Temperature dependence of pre-edge XAS peaks]
{Temperature dependence of the pre-edge structure in the Ti K-edge
x-ray absorption spectrum of rutile}%

\author{O. Durmeyer, E. Beaurepaire, J.-P. Kappler}
\address{%
IPCMS, UMR 7504 CNRS-ULP, 23 rue de Loess, B.P. 43 Strasbourg Cedex 2, 
France 
}%
\author{Ch. Brouder}
\address{%
Institut de Min\'eralogie et de Physique des Milieux Condens\'es,
CNRS UMR 7590, Universit\'es Paris 6 et 7, IPGP, 140 rue de Lourmel,
75015 Paris, France
}%
\author{F. Baudelet}
\address{%
Synchrotron SOLEIL,
L'Orme des Merisiers,
Saint-Aubin - BP 48,
F-91192 Gif-sur-Yvette cedex, France.
}%

\ead{christian.brouder@impmc.jussieu.fr}
\begin{abstract}
The temperature dependence of the pre-edge features
in x-ray absorption spectroscopy is reviewed.
Then, the temperature dependence of the pre-edge structure
at the K-edge of titanium in rutile TiO$_2$ is measured
at low and room temperature. The first two peaks grow
with temperature. The fact that these two peaks also correspond to 
electric quadrupole transitions is explained by
a recently proposed theory.
\end{abstract}
\today

\pacs{78.70.Dm, 63.70.+h}
\maketitle

\section{Introduction}

Pre-edge peaks often arise at the K-edge of transition metal
elements. This pre-edge structure is sensitive to the metal
valence, to the symmetry of its surroundings and to
the atomic species of the neighbours
(see \cite{Yamamoto-08} for a recent review).
As a consequence, the measurement and analysis
of the pre-edge peaks are widely used in
earth sciences~\cite{Chalmin-09},
biology~\cite{Arfaoui-07},
chemistry~\cite{Getty-08}
and physics~\cite{Gougoussis-09,Vedrinskii-09}.

Because of their practical importance,
pre-edge features have to be well understood
and they were the object of detailed
theoretical works using various approaches:
multiplets~\cite{Arrio-00,Groot-09},
Bethe-Salpeter equation~\cite{Shirley-04},
multiple-scattering~\cite{Wu-03},
pseudopotential theory~\cite{Gougoussis-09,Juhin-08}.
Vedrinskii and his group were particularly active to extract
information from the pre-edge 
structure~\cite{Vedrinskii-97,Vedrinskii-98,Vedrinskii-06}.

In section 2, we make a short review of the literature 
to show that the temperature dependence of pre-edge peaks 
is not a rare property of x-ray absorption spectra.
However, this dependence is usually attributed to
static off-center displacements or to phase transitions.
Therefore, our preliminary investigation~\cite{Durmeyer} 
showing a temperature dependence of the pre-edge peaks
at the titanium K-edge in TiO$_2$ (rutile)
came out as a surprise because the pre-edge variation was
observed in a temperature range where no phase transition
occurs and where many high-precision structural 
studies~\cite{Burdett-87,Howard-91,Sakata-93,Sabine-95,Kumazawa-95}
indicate that no off-center atomic displacement takes place.
Soft modes have indeed be reported~\cite{Jiang-03} but
the calculated phonon spectrum shows excellent agreement
with experiment and no imaginary mode is 
present~\cite{Montanari-04,Rignanese-05,Rignanese-05-Err,Sikora-05}
when the proper functionals are used~\cite{Montanari-02}.

Thus, we carried out detailed experiments to confirm and
analyze this temperature dependence. The results of these
experiments are presented in section 3. 
Section 4 describes why such a temperature dependence is
a priori surprising and sketches a theoretical interpration
that enables us to understand why the temperature
dependence is restricted to the first two peaks and
why no energy shift is observed.
A conclusion summarizes our results and provides possible extensions
of this work.

\section{A short review}
In this section, we present a short and non-exhaustive review of
the temperature dependence of pre-edge peaks.

As far as we know, such a temperature dependence
 was first observed by Durmeyer 
and coll.~\cite{Durmeyer} at the K-edge of
titanium in TiO$_2$ (rutile), Li$_{4/3}$Ti$_{5/3}$O$_4$ and
LiTi$_2$O$_4$. It was subsequently measured at
the titanium K-edge of a several perovskite crystals:
PbTiO$_3$~\cite{Ravel-93,Ravel-97,Vedrinskii-97,Sato-05,Hashimoto-07}, 
SrTiO$_3$~\cite{Nozawa,Hashimoto-07}, 
BaTiO$_3$~\cite{Hashimoto-07}, 
CaTiO$_3$~\cite{Hashimoto-07}.

A similar temperature dependence was observed 
at other edges in perovskite crystals:
at the niobium K-edge of
KNbO$_3$~\cite{Shuvaeva-99,Shuvaeva-03},
NaNbO$_3$~\cite{Shuvaeva-01},
PbIn$_{1/2}$Nb$_{1/2}$O$_3$~\cite{Shuvaeva-04},
at the zirconium K-edge of
PbZrO$_3$~\cite{Vedrinskii-06},  
PbZr$_{0.515}$Ti$_{0.485}$O$_3$~\cite{Vedrinskii-06},  
BaZrO$_3$~\cite{Vedrinskii-06}, at
the K-edge of Mn in La$_{1-x}$Ca$_x$MnO$_3$~\cite{Qian-00,Bridges-00}
and at the K-edge of Fe in 
in La$_{0.8}$Sr$_{0.2}$FeO$_3$ and
La$_{0.7}$Sr$_{0.2}$FeO$_3$~\cite{Deb-06}.
In most cases, the temperature effect was intepreted 
in terms of a phase transition or of a static off-center
atomic displacement due to the
presence of very soft modes in the crystal.

However, the effect is not restricted to the perovskite
structure. Apart from the results of Durmeyer and coll.~\cite{Durmeyer},
it was observed at the K-edge of titanium in 
TiO$_2$ and Mg$_2$TiO$_4$~\cite{Hashimoto-07},
at the L-edges of La in Sr-doped La$_2$CuO$_4$~\cite{Hidaka-03}
and at the K-edge of V in VO$_2$~\cite{Poumellec-94}.
A temperature dependence of XANES spectra was also
observed at the K-edge of oxygen in
water~\cite{Wernet-04} and in 
doped LaMnO$_3$~\cite{Mannella-05,Tsai-09}.
Finally, the Mahan-Nozi{\`e}res-Dominics singularity
can also give rise to a temperature dependence of
the x-ray absorption spectra of metals
(see Ref.~\cite{Ohtaka-90} for a review).

We come now to our experimental temperature dependence
at the K-edge of titanium in rutile.

\section{Experiment}

The x-ray absorption experiments were performed at the D11
(energy dispersive) and at the EXAFSII stations of the DCI storage ring of the 
Laboratoire pour l'Utilisation
du Rayonnement Synchrotron in Orsay (France).

A rutile single-crystal plate (9~mm x 4~mm x 50~$\mu$m) was measured 
at the D11 station in the transmission mode.
The crystal plate was placed inside a liquid-helium cryostat operating 
between 4.2~K and 300~K.
Measurements were carried out for two orientations, 
with the (110) face of the crystal perpendicular to the x-ray beam
and the $c$-axis either parallel or perpendicular to the linear polarization 
vector of the beam.
The polychromator consisted in a curved Si(111) crystal focusing
the X-ray beam at the center of the cryostat sample holder. Higher harmonics 
were rejected by a SiO$_2$ plane mirror. The x-ray intensity was measured by 
a photodiode array
detector. Each spectrum was obtained as a result of four measurements:
$I_0$ (without sample and with the beam),
$I_{0\mathrm{black}}$ (without sample and without beam),
$I$ (with sample and with the beam),
$I_{\mathrm{black}}$ (with sample and without beam).
The absorption spectrum was then obtained from the formula
$\sigma=\log(I_0-I_{0\mathrm{black}})-\log(I-I_{\mathrm{black}})$.
The x-ray energy corresponding to each detector pixel was determined by
comparing the spectra with a spectrum measured on a two-crystal monochromator beamline.
The energy resolution was typically 0.8 eV.

Our preliminary study~\cite{Durmeyer} showed us that the pre-edge structure
could exhibit a low signal to noise (S/N) ratio when the crystal thickness was
optimized for the edge-jump. Therefore, we optimized the crystal thickness
for the pre-edge structure and cut an approximately 50~$\mu$m thick 
crystal plate.
As a consequence, we obtained excellent spectra in the pre-edge region but 
the XANES spectra after the edge had a rather low S/N  ratio
and, for each polarization direction, we normalized the spectrum at the
inflection point of the absorption edge instead of at the edge jump.

To check the validity of this procedure, we carried out additional experiments
at the EXAFSII station. The experimental equipment consisted of a two-crystal
Si(311) monochromator, an ionisation chamber to measure the incident beam
and an electron-yield detector.
We measured a bulk rutile single crystal with the (110) face perpendicular
to the x-ray beam and with the $c$-axis of the crystal
either parallel or perpendicular to the x-ray polarization vector.
The S/N ratio of the pre-edge region was
comparable to that of the Ti K-edge spectra of rutile  measured 
on the same beamline in similar conditions~\cite{BrouderRome,Poumellec-91}.
The experimental spectra were normalized by the standard procedure 
and, as in our previous work~\cite{Durmeyer}, the temperature dependence was found 
to be negligible except in the pre-edge region.
Moreover, the observed spectra and temperature dependence agreed well with the transmission 
experiments of the D11 station. 
In the present paper we show only the results of the transmission experiments because
of their better S/N ratio.

\section{Experimental results}
\begin{figure}
\includegraphics[width=8.0cm]{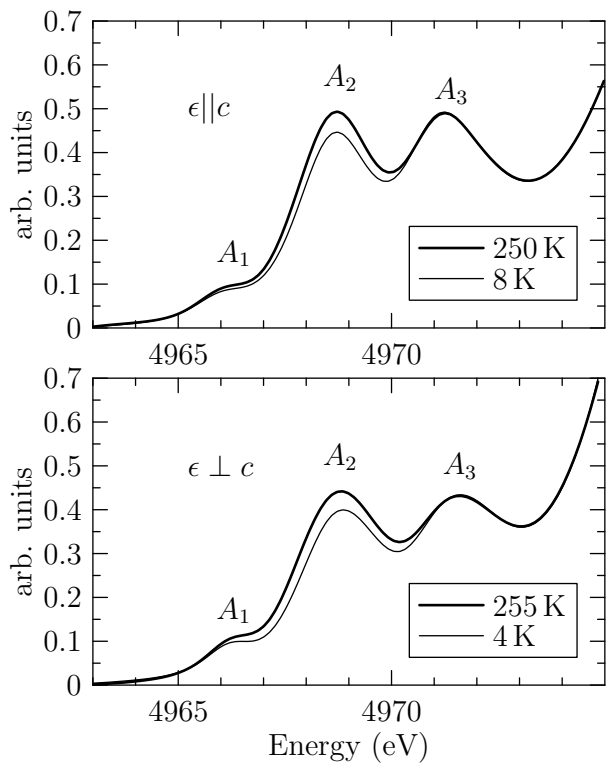}
\caption{Temperature variation of the pre-edge peaks
at the K-edge of titanium in rutile, with the
x-ray polarization vector parallel and perpendicular
to the $c$-axis of the crystal\label{figtic}.}
\end{figure}

Figure~\ref{figtic}  shows the pre-edge features of rutile recorded at
different temperatures with the polarization vector perpendicular
and parallel to the crystal $c$-axis. 
A decrease  of the first two peaks A$_1$ and A$_2$  is observed
at low temperature, whereas the third peak A$_3$ does not show any
significant variation. 
It is important to notice that temperature induces a change
in the intensity but not in the energy position of the peaks.

The physical origin of the pre-edge peaks of titanium
in rutile is well known~\cite{Joly-99}.
Peaks A$_1$ and A$_2$ correspond to electric quadrupole
transitions towards $3d$ states of titanium with $t_{2g}$ and $e_g$ 
symmetry, respectively.
Therefore, the peaks that vary with temperature are also the
peaks corresponding to quadrupole transitions.

\begin{figure}
\includegraphics[width=8.0cm]{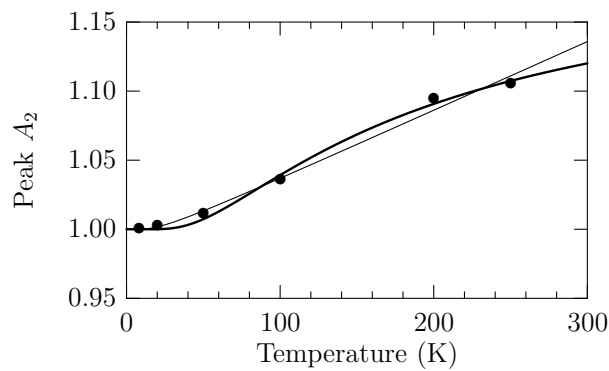}
\caption{Relative intensity of peak $A_2$ as a function
of temperature for $\epsilon\,||\,c$, normalized to 1
at 8K. Dots: experimental results;
thick solid line: fit to the function
  $a (1+\e^{-\theta/T}) +(1-a)$;
thin solid line: fit to the function
  $a \cosh(\theta/2T) +(1-a)$.\label{figta2}}
\end{figure}

In the optical range, the effect of temperature is
usually described by a simple model developed
by Holmes and McClure~\cite{Holmes-57,McClure-59,Triest-99},
in which the intensity of the vibronic peak varies as  $1+e^{-\theta/T}$,
where $\theta$ is the energy of the first vibrational level.
Figure ~\ref{figta2} shows the variation of the $A_2$ peak
with temperature, fitted to the
function $a(1+e^{-\theta/T}) + (1-a)$,
where $a(1+e^{-\theta/T})$ represents the fraction of the
$A_2$ peak that is purely vibrational and 
$1-a$ the fraction 
that is due to electric quadrupole transitions
(and to the possible tail of the electric dipole peak $A_3$).

The result of the fit is $a=0.21$
and $\theta=168~K~\pm 10~K$. Note that the value of $\theta$
compares favorably to the energy of the first odd
vibrational level at the $\Gamma$ point
obtained by \emph{ab initio} calculations
(168~K~\cite{Lee-94}, 150~K~\cite{Montanari-02},
169~K~\cite{Sikora-05} or 181~K~\cite{Rignanese-05})
or by neutron scattering 163~K~\cite{Traylor-71}.
However, the simplicity of the Holmes and McClure model
implies that the quality of this agreement is probably
fortuitous.
Indeed, an alternative single-mode model is sometimes 
used~\cite{Pisarev-69,Cieslak-80,McDonald-86,Taran-94}, for which
the temperature dependence is
$\coth{\theta/2T}$. For this second model the fit gives
$\theta=58~K~\pm 5~K$ and $a=0.014$. 

\section{Interpretation}
It remains to understand why only the first two peaks
vary with temperature while the rest of the XANES spectrum remains
constant. We first describe the arguments that are usually
given to explain the \emph{absence} of temperature dependence of
XANES spectra. Then, we show why, in some circumstances,
this independence can be broken.

\subsection{The temperature independence of XANES spectra}
There are many reasons to believe that the pre-edge features
of x-ray absorption spectra do \emph{not} depend on temperature
in the absence of structural transition.
The first reason comes from the temperature dependence of
the EXAFS part of x-ray absorption spectra, which is
represented by a Debye-Waller
factor $\ee^{-2k^2\sigma}$ in the EXAFS formula.
The Debye-Waller factor 
accurately describes the temperature dependence of XAS in crystals,
although it has to 
be supplemented with higher order cumulants in disordered materials.
Moreover, it is well understood because it can be calculated 
\emph{ab initio} with a good agreement with 
experiment~\cite{Poiarkova,Chun-04,Vaccari-05,Dimakis,Vila-07}.
If we use this factor to describes the temperature
dependence near the edge, we must take an energy very
close to the Fermi energy, so that $k$ is very small
and the factor is close to unity.

Of course, near the edge, the effect of temperature
is not supposed to be described by a Debye-Waller factor
and we must use a more sophisticated approach.
Natoli's rule~\cite{Natoli-83} gives good results
near the edge. However, it describes an energy
shift through the equation $k R=$Constant,
and we do not observe any energy shift.
More elaborate theoretical analyses were carried out.
Brouder and Goulon~\cite{BrouderGoulon,BrouderLie} used Lie group theory 
to describe the influence of a displacement on the 
multiple-scattering operator. However, the
temperature evoluation is expected to be small near the edge,
essentially because of Natoli's rule.
Poiarkova and Rehr~\cite{Poiarkova} extended the Debye-Waller factor to
multiple-scattering paths. Their formalism is not really 
valid in the pre-edge region, but if we try to extend it 
we find a very small temperature dependence because of the
presence of the $k^2$ factor in the exponent.
Fujikawa~\cite{Fujikawa-96,Fujikawa-96-JPSJ}
used Schwinger's technique to calculate the effect of the 
Franck-Condon factors on XAFS. He concluded that
this effect was not important.
In a later work~\cite{Fujikawa-99}, he investigated the
effect of temperature through the 
Keldysh approach to nonequilibrium systems.
He found that thermal vibrations could be represented
by a convolution with the phonon spectral function. 
His result is valid in the pre-edge region but
leads to a small temperature effect. Moreover,
the convolution should give rise to a broadening
of the pre-edge peaks with temperature.
Again, this is not compatible with our experimental results.
A further elaboration of his approach~\cite{Arai-XAFS13-phonon} lead to 
similar results.

We can try to take vibrations into account by 
coming back to the Born-Oppenheimer approximation and
writing the initial and final wavefunctions as a product
of a vibrational and an electronic function.
However, this approach looks like a dead end if we consider the work
by M\"ader and Baroni~\cite{Mader} who showed that, at the K-edge
of carbon, the vibrations in the final state are
strongly anharmonic and are deeply affected by
the presence of the core-hole.
Therefore, we are not allowed to consider the vibrations
as similar in the initial and final states and we cannot
use the harmonic approximation.

Ankudinov and Rehr~\cite{Ankudinov-XAFS12} brought 
some hope by showing that the S K-edge spectrum of 
SF$_6$ is closer to experiment when the atomic positions
are slightly shifted with respect to the equilibrium position.
But, as can be seen in their figure, atomic displacements
shift the position of the main lines and this shift is not
experimentally observed.

\subsection{The temperature dependence of XANES spectra}
We recently proposed a model that enables us to
understand the observed temperature dependence~\cite{BC-09}.
Although a detailed account of this model would be
beyond the scope of the present paper, we can give
a physical description of the underlying physics.

We start from the Born-Oppenheimer approximation where
the wavefunctions of the electron+nuclei system is 
the product of a vibrational function
by a solution of the Schr\"odinger equation for clamped
nuclei. The energy of these wavefunctions do not
depend on the position of the nuclei (as the 
eigenvalues of the Schr\"odinger equation for an
electron in a potential do not
depend on position). The transitions are made between
these wavefunctions. If we assume that the vibrational
energies are small with respect to experimental resolution,
we can sum over the final state vibrational functions
and we obtain an average over the vibrational function
of the initial state of transitions for which the transition 
energy do not depend on the atomic positions.
This explains why the peak positions do not
move while they move if we calculate the spectrum of
a distorted structure.

The second step of the model consists in making a different
approximation for the initial and the final states.
The initial state is taken to be the core state centered
at the position specified by the vibrational wave function.
For the final state, we make the \emph{crude} Born-Oppenheimer 
approximation, where the electronic wavefunction is assumed
independent of the position of the absorbing atom.
Then, the cross-section boils down to an average of the 
x-ray absorption spectra for a shifted core wavefunction
(with fixed energies).
What happens next can be sketched by an oversimplified 
model of the shifted core wavefunction.
We assume that the displacement $\bfR$ is small compared to 
the electronic variable $r$
and we obtain, to first order in $\bfR$ and for a spherical
core state $\phi_0(r)$, the shifted function
\begin{eqnarray}
\phi_0(|\bfr-\bfR|) &\simeq& \phi_0(r) - \frac{\bfr\cdot\bfR}{r}
\phi_0'(r).
\label{faux}
\end{eqnarray}
When multiplied by $\epsilon\cdot\bfr$, the 
additional term gives us a factor
$\epsilon\cdot\bfr \bfR\cdot\bfr$ that
can be transformed into
the sum of a monopole term proportional to 
$(\epsilon\cdot\bfR) r^2$ and a quadrupole term.
The monopole term gives rise to transitions towards $s$ states,
the quadrupole term to transition towards $d$ states.
The transition towards $s$ states are observed 
at the aluminium or silicon K-edge~\cite{BC-09}, the transitions
towards $d$ states are observed at the K-edge
of transition metals because of the presence of
a strong density of $d$ states.
This explains why the temperature variation occurs at
the position of the quadrupole peaks.
Finally, the fact the temperature-dependent pre-edge
peaks grow with temperature is due to the corresponding increase in
thermal vibration amplitudes.

Of course, eq.~(\ref{faux}) is not sufficient because
the integration over $\bfr$ includes also a region where
$r < R$. The full theory~\cite{BC-09} is more complex
but the physical idea is the same.

\section{Conclusion}
In this paper, we have presented the
temperature-dependence of pre-edge features at the
K-edge of titanium in rutile. This temperature dependence
is not due to a phase transition or to a static
distortion of the titanium site.

The temperature dependence changes only the intensities
of the peaks and not their positions. Moreover, the
peaks that vary with temperature are the electric quadrupole
peaks of the spectrum. An explanation of this behavior
was given in terms of the dynamic displacement of the
absorbing atom.

Two conditions turn out to be crucial to observe 
temperature-dependent pre-edge peaks at the K-edge :
(i) a large density of $d$ states below the $p$ states,
so that the transitions to final $d$ states are
significant and visible;
(ii) the existence of low-energy vibrational modes,
so that the temperature effect can be observed at
reasonable temperatures. 
Both of these conditions are satisfied in rutile
and in perovskites containing transition metals.
In that case, the temperature dependence provides
information on the local vibrations around the 
absorbing atom. This can be particularly useful
to investigate the vibrations of transition metal impurities.

\ack
We thank Delphine Cabaret for very constructive comments.

\section*{References}


%

\end{document}